**Dependence of Empirical Fundamental Diagram on Spatial-Temporal Traffic Patterns Features**

Boris S. Kerner

DaimlerChrysler AG, RIC/TS, HPC: T729
D-70546 Stuttgart, Germany
Phone: + 49 711 17 41453, Fax: +49 711 17 41242
boris.kerner@daimlerchrysler.com

**Abstract**

It is shown that the branch for congested traffic in the empirical fundamental diagram strong depends both on the type of the congested pattern at a freeway bottleneck and on the freeway location where the fundamental diagram is measured. The type of the pattern at the same bottleneck can depend on traffic demand. Thus, at the given freeway location a qualitative form of the empirical fundamental diagram can strong depend on traffic demand also. At freeway locations where the average speed in synchronized flow is higher the branch of the fundamental diagram for congested traffic has a *positive slope* in the flow-density plane as well the slope of the branch for free flow. The average positive slope of the branch for congested traffic decreases in the upstream direction when the average synchronized flow speed inside the congested pattern decreases. The branch for congested traffic with a positive slope at a downstream location can transform in the upstream direction in the branch for congested traffic which has a *maximum* in the flow-density plane: At lower density this branch has a positive and at higher density it has negative slope. At freeway locations, where the average speed in synchronized flow further decreases the branch for congested traffic more and more tends to the line J which represents the stationary motion of the downstream front of a wide moving jam in the flow-density plane: The line J is an asymptote for the branch of congested traffic of empirical fundamental diagrams higher densities for those freeway location where wide moving jams occur and propagate.

# INTRODUCTION

The empirical fundamental diagram (the flow-density relationship) is a very important average characteristic of traffic flow: (i) This characteristic is used in a lot of engineering applications and (ii) the empirical fundamental diagram is an empirical basis for a theoretical fundamental diagram. The latter is used in a huge number of traffic flow theories and models within the fundamental diagram approach. Corresponding to a lot of empirical studies made in different countries, the empirical fundamental diagram consists of two isolated branches (curves): The branch for free flow and the branch for congested traffic (e.g., *(1)-(6)*).

At freeway bottlenecks dependent on the bottleneck type and traffic demand qualitative different *spatial-temporal* congested patterns can occur. Recently, empirical spatial-temporal features of congested patterns have been found and their classification have been made *(7)-(9)*. Empirical spatial-temporal congested pattern features of *(7)-(9)* confirm main hypotheses of author's three-phase traffic theory *(10), (11)*. Corresponding to this theory, there are two traffic phases in congested traffic: "synchronized flow" and "wide moving jam". In *(7), (8)*, it has been found that there are different synchronized flow patterns (SP) [the widening SP (WSP), the localized SP (LSP) and the moving SP (MSP)] and general patterns (GP) which can occur at an isolated bottleneck. SP consists of synchronized flow only. In GP synchronized flow is formed upstream of the bottleneck and wide moving jams spontaneously emerge in that synchronized flow, i.e., GP consists of the both traffic phases in congested traffic, "synchronized flow" and "wide moving jam". If two or more bottlenecks are close to one another on a freeway then expanded congested patterns (EP) can appear. In EP, synchronized flow covers two or more bottlenecks.

Due to a spatial dependence of mean values of traffic variables along of these patterns, we may find very different mean characteristics of these traffic variables at different freeway locations upstream of the bottleneck. Thus, we may expect that if average characteristics of freeway traffic are studied, then these characteristics should depend on both the pattern type and on a freeway location. This supposition concerns also the flow-density relationship, i.e., the empirical fundamental diagram. In this paper, we will show that this supposition is true: The empirical fundamental diagram indeed strong depends both on the type of the congested pattern and on the freeway location where the fundamental diagram is measured. To study this feature of the empirical fundamental diagram, we will make a comparison of data in the flow-density plane related to synchronized flow inside a congested pattern with the line J. The sense of this comparison can be clear if we recall the definition of the line J *(7), (12)*.

## Definition and determination of the line J

The line J represents the stationary propagation of the downstream front of a wide moving jam in the flow-density plane. The line J is determined by the line slope and by one co-ordinate in the flow-density plane. This slope is equal to the velocity of the downstream jam front, $v_g$. The co-ordinate on the line J is given by the average traffic flow variables related to the jam outflow (the flow rate and the density in the jam outflow). The maximum of the flow rate in this jam outflow, $q_{out}$ is reached if free flow is formed in the wide moving jam outflow.

Let us inside either GP or EP a sequence of wide moving jams is formed. We suggest that there is a detector which measures traffic variables in this jam sequences. Then, during time intervals when these jams are not at the detector, the average traffic variables measured at the detector are related to the outflow from a wide moving jam in this sequence which is downstream of the detector location.

Therefore, corresponding to the line J definition, the average synchronized flow variables for this jam outflow should give a point in the flow-density plane which lies on the line J, if no on- and off-ramps are between the detector and the jam *(7), (13)*. In other words, averaged measured data in synchronized flows which are related to different wide moving jam outflows should give different points. All these points are on the line J in the flow-density plane. In particular, if a moving jam just transforms into a wide moving jam then the measured averaged data related to the jam outflow should tend to the line J in the flow-density plane *(12)*. This empirical result is linked to the above conclusion that empirical averaged characteristics of the flow in the wide moving jam outflow are related to points in the flow-density plane which lie on the line

J *(7), (13)*. This conclusion of a study of initial 1-min data in *(7), (13)* is confirmed by an analysis of single vehicle data in *(14)*.

When a moving jam is not a wide moving jam or there is no moving jams in synchronized flow at all, then averaged data related to synchronized flow should not necessary lie on the line J *(7), (9)*. It occurs that dependent on the vehicle speed in a synchronized flow the average points related to this synchronized flow show qualitatively different flow-density relationships.

We will also find that these flow-density relationships can possess a qualitative different shape in comparison with the line J. However, let us the average speed in synchronized flow is a monotonous decreasing function of spatial co-ordinate inside a congested pattern. At freeway locations where the synchronized flow speed is relative high, the fundamental diagram is qualitatively different from the line J. Due to the spatial decrease in the speed, wide moving jams emerge in the synchronized flow at the freeway locations where the speed is low enough. Then there is an *asymptotic* behavior for all these different flow-density relationships as functions of the spatial co-ordinate: This asymptotic is given by the line J. This may explain the importance of the comparison of the part of the empirical fundamental diagram for congested traffic with the line J in the flow-density plane.

**MAIN STAGES OF FUNDAMENTAL DIAGRAM SPATIAL EVOLUTION**

To study a possible spatial dependence of empirical fundamental diagram, we will first make a comparison of data in the flow-density plane which are related to an example of WSP at the bottleneck due to the off-ramp from *(9)* (Fig. 1) with the line J (Figs. 2-4).

First it should be recalled that a congested pattern can be considered as a congested pattern at the isolated bottleneck due to the off-ramp only up to the detectors D17: The upstream detectors D16 are already related to the next upstream effective bottleneck on the freeway section (Fig. 1 (a)). Due to the upstream propagation of synchronized flow of the initial WSP at the off-ramp, a EP is formed. The synchronized flow in the EP covers the upstream bottleneck at the on-ramp D16. The part of EP which is related to D17-D24, where no wide moving jams occur, can be considered as WSP upstream of the bottleneck due to the off-ramp D25-off (Fig. 1(b, c)).

It can be seen from 1-min data for the left (passing) lane of the freeway (where tracks must not move) for a part of the WSP (D19-D21) that empirical points for synchronized flow are deviate qualitatively from the line J (Fig. 2). The higher the density $\rho$, the higher is a mistake in the calculation of the density through the formula $\rho = q / v$ where the average speed $v$ and the flow rate $q$ are measured traffic variables. For this reason we will not use those densities $\rho$ which are higher than 70 vehicles/km.

To find the empirical fundamental diagrams at different freeway locations, we divide the density-axis in the flow-density plane on small density ranges (3 vehicles/km) and then we average 1-min data to only one point for each of these density ranges. Then, an approximation of these averaged free flow states and states of congested traffic (synchronized flow states plus states which are related to the fronts of wide moving jams) gives us the flow-density relationships, i.e., the empirical fundamental diagrams for the left freeway lane (Fig. 3) and the diagrams related to data averaged over all three freeway lanes (Fig. 4).

Downstream of WSP, at the detectors D24 only free flow is formed: There is only the branch for free flow F (Figs. 3 and 4, D24). At locations of the detectors D21-D17 we first find the well-known result (e.g., *(1)-(5)*) that there are two branches of the empirical fundamental diagram: (i) The branch F for free flow and (ii) the branch C for congested traffic (Figs. 3 and 4, D21-D17). However, considering the WSP in the upstream direction from the location of the detectors D21 to the location of the detectors D17, we can see that the branch for congested traffic C changes *qualitative* at different freeway locations inside the congested spatial-temporal pattern (Figs. 3 and 4). In particular, we find the following.

The points of synchronized flow at the location D21 (curve C) has a positive slope in the flow-density plane. This slope should approximately be related to the velocity of very small amplitude perturbations in this synchronized flow. The positive slope of the curve C related the synchronized flow states remains also for the locations upstream at the detectors D20 up to the detectors D19. However, the positive slope of the branch C decreases in the upstream direction.

At the location D18 a new effect occurs: At lower density the branch C has a positive slope. However for higher density the branch C has a negative slope, i.e., the branch C as a function of the density has a maximum point.

At the location D17 narrow moving jams emerge in synchronized flow. Apparently for this reason the curve C is above the line J (see explanations in *(10)*) and it has a negative slope. This slope is a little bit more negative than the slope of the line J. This may correspond to the conclusion *(10)* that narrow moving jams have a more negative velocity than the characteristic velocity $v_g$ for the downstream front of a wide moving jam. Upstream of D17 due to the upstream bottleneck at the on-ramp D15-on as it has above been mentioned the congested pattern is a EP, i.e., the pattern may not be considered as a congested pattern at the isolated bottleneck any more.

At the location D16 the dynamics of synchronized flow in this EP should essentially be determined by the upstream bottleneck due to the on-ramp D15-on (Fig. 1(a)). In the case, some narrow moving jams transform into wide moving jams. As a result, the branch C at D16 almost coincides with the line J.

The above consideration is related to the left freeway lane. Recall that in German three-lane (in one direction) freeways the left lane is the passing lane. This lane must not be used by tracks and other long vehicles. "Aggressive" drivers usually prefer this passing lane. In contrast, on the middle and the right lanes long vehicles may move. However, this difference in vehicle's parameters (and probably in driver's characteristics) do not change *qualitative* results about spatial dependence of the fundamental diagram for the left lane. However, there are some quantitative changes in characteristics of averaged synchronized flow states over all three lanes (Fig. 4) are sometimes different from that for the left lane (Fig. 3).

Qualitative results about the dependence of the branch C on the spatial co-ordinate which have been considered for the left freeway lane do not depend on the different vehicle's parameters and driver's characteristics of real traffic flow. This can be seen if the empirical fundamental diagrams for the left lane (Fig. 3) and for data averaged over all three freeway lanes (Fig. 4) are compared. However, there are some quantitative changes both in the branch of free flow F and in the branch of congested traffic C.

In particular, we can mention the following differences:

(1) The maximum in the branch for congested traffic C at the location D18 is a smooth one for the data averaged over all three lanes (Fig. 4) in comparison with the relative sharp maximum for the left freeway lane (Fig. 3).

(2) Whereas for the left lane the maximum flow rate in free flow has been higher than the maximum flow rate in the wide moving jam outflow $q_{out}$ at all detectors (Fig. 3), we find this result only at the locations D16-D18 for the data averaged over all three lanes (Fig. 4). At locations D21-19 the maximum flow rate in free flow is even lower than $q_{out}$. This may be explained by two reasons. First, some part of the vehicles which want to leave the main road to the off-ramp D25-off chose the left lane already some kilometers upstream of the off-ramp. Second, this is a strong asymmetry in traffic rules on different freeway lanes on German three lane (in one direction) freeways. Note that a difference between flow-density relationships measured at one location for different freeway lanes is also observed on US-freeways (see review by Banks *(9)*).

(3) The branch C for congested traffic at the location D16 are close but above the line J for the data averaged over all three lanes (Fig. 4). This is probably because at the location D16 some of moving jams do not possess the characteristic velocity $v_g$ of wide moving jams for all three freeway lanes, i.e., they are still narrow moving jams on the middle and/or right lanes.

This consideration allows us to distinguish the following main stages of the qualitative dependence of the empirical fundamental diagram on the freeway location (the spatial co-ordinate):

(i) At freeway locations where the average speed in synchronized flow is higher we find that the branch of the fundamental diagram for congested traffic C has a positive slope in the flow-density plane as well the branch for free flow F. The average slope of the branch for congested traffic decreases in the upstream direction when the average synchronized flow speed decreases.

(ii) At upstream freeway locations where the average speed in synchronized flow is further decreasing we find that the curve for congested traffic C has a *maximum* in the flow-density plane: At lower density the branch C has a positive and at higher density it has negative slope.

(iii) At freeway locations, where the average speed in synchronized flow further decreases, moving jams emerge in that synchronized flow. In this case, the branch for congested traffic C at higher density more and more tends to the line J.

(iv) Finally, at locations where wide moving jams occur in synchronized flow the branch for congested traffic at higher density is related to the line J in the flow-density plane.

Already from these results we see that the line J plays a role of an asymptote for the branch of congested traffic C of the empirical fundamental diagram: In the region of a congested pattern where wide moving jams are realized the branch C at higher density almost coincides with the line J. This result is linked with the physical sense and the definition of the line J: Each average point of measured data which is related to the wide moving jam outflow should lie on the line J in the flow-density plane *(7)*.

## ASYMPTOTIC BEHAVIOR OF FLOW-DENSITY RELATIONSHIPS

### Empirical fundamental diagrams for GP under strong congestion

The asymptotic behavior of the branch C for congested traffic of empirical fundamental diagrams is essentially important for GP and EP where wide moving jams occur. Let us first compare measured data at different locations inside a GP under strong congestion (Fig. 5 (a, b)) *(7)*. Because the results for different GPs which occur at different days are very similar in all cases we will restrict the analysis of only one GP.

At the location D7, i.e., downstream of the bottleneck due to the on-ramp D6-on only free flow occurs (Fig. 5 (c)). At the bottleneck a GP emerges: At the bottleneck due to the on-ramp (D6 in Fig. 5 (a)) (the bottleneck is marked $B_3$ in Fig. 5 (b)) synchronized flow is realized. Upstream of the bottleneck at the locations D5-D4 the pinch region occurs in the synchronized flow of GP where narrow moving jams emerge *(7)*. Some of these narrow moving jams transform into wide moving jams at the locations D3-D2.

The flow-density relationships (Fig. 6) show the same common general qualitative results of the dependence of empirical fundamental diagrams of the spatial co-ordinate which have been considered for WSP above. In particular, at the freeway location D6 where the average speed in synchronized flow is higher we find that the branch of the fundamental diagram for congested traffic C has a positive slope in the flow-density plane as well the branch for free flow F.

At upstream freeway locations D5-D4 where the average speed in synchronized flow is further decreasing we find that the branch for congested traffic C has a maximum in the flow-density plane.

At freeway locations, D3-D1 where wide moving jams occur the branch C for congested traffic for higher density is related to the line J in the flow-density plane.

However, we also find some new peculiarities. First, recall that for the WSP there is a long section of the freeway between locations D21-D19 where the branch of the fundamental diagram for congested traffic C has a positive slope in the flow-density plane. In contrast to this behavior, in the case of GP under strong congestion there is no such long freeway section: Already at the location D5 just upstream of the bottleneck the branch for congested traffic C has a maximum in the flow-density plane: At lower density the slope of this branch is positive and at higher density the slope is negative. This is linked to the pinch effect in synchronized flow of GP. In the pinch region of GP narrow moving jams emerge propagating upstream. For this reason, at higher densities the slope of the flow-density relationship is negative. This slope is more negative than the slope of the line J. This means that narrow moving jams in the pinch region have a more negative velocity than the velocity of the downstream front of a wide moving jam. Besides, this part of the branch for congested traffic at higher density is above the line J. This is related to the result *(10)* that states of synchronized flow in the pinch region at higher density are above the line J in the flow-density plane (D5, Fig. 6).

In the region of wide moving jams of the GP (the locations D3-D1) the branch C for congested traffic of the fundamental diagrams at higher densities lies on the line J. Thus, the line J is an asymptotic for the fundamental diagrams of traffic flow at higher densities for those freeway locations where the region of wide moving jams is formed inside GP.

This and other conclusions made for the left freeway lane are qualitatively valid for the other lanes as it can be seen from Fig. 7 where the data are averaged over all three freeway lanes. There are only some quantitative peculiarities in the latter case which as well for WSP are probably linked with the asymmetry between different lanes and different vehicle's parameters and driver's characteristics. In particular, the difference between the maximum flow rate in free flow and the maximum flow rate in the wide moving jam outflow $q_{out}$ is at some locations lower than for the left lane.

Empirical fundamental diagrams for GP under weak congestion

If we now study measured data at different locations inside a GP under weak congestion (see Fig. 20 (a) in *(7)*) then we find that the dependence of fundamental diagrams on different freeway locations shows intermediate features between the ones for the WSP (Fig. 3) and for the GP under strong congestion (Fig. 6).

At the location D27, i.e., downstream of the bottleneck due to the off-ramp D25-off (Fig. 1 (a)) only free flow occurs (Fig. 8 (a)). At the bottleneck the GP emerges: Upstream of the bottleneck at the locations D24-D23 synchronized flow occurs. In the synchronized flow of GP, wide moving jams emerge (D22-D19).

At freeway locations D24-D23 the branch of the fundamental diagram for congested traffic C has a positive slope in the flow-density plane as well the branch for free flow F (Figs. 8 (a) and 9). The line J is an asymptotic for the fundamental diagrams of traffic flow at higher densities for those freeway locations D22-D19 where the region of wide moving jams is forming inside GP. The branches for congested traffic at the freeway locations D22-D19 possess a maximum point: At lower density they have a positive slope.

One the one hand, it can be seen that in contrast to the GP under strong congestion, for the GP under weak congestion similar to that for the WSP there is long enough section of the freeway (D24-D23) where the branch of the fundamental diagram for congested traffic C has a positive slope in the flow-density plane (Figs. 8 (a) and 9). On the other hand, as well in the GP under strong congestion (Fig. 6) there is also a long freeway section (D22-D19) where wide moving jams occur in the GP and therefore the line J is a good asymptotic for the fundamental diagrams of traffic flow at higher densities.

**Empirical fundamental diagrams for EP**

The line J as the asymptotic for the fundamental diagrams of traffic flow at higher densities is a very often result for the locations inside EP. Recall that EP occurs if there are at least two effective bottlenecks, the downstream and upstream bottleneck which are close to one another. Besides, at the downstream bottleneck a congested pattern should be formed.

EP occurs for example due to the upstream propagation of synchronized flow of the initial GP under weak congestion considered above. In this case, the downstream bottleneck is due to the off-ramp D25-off and the upstream bottleneck is due to the on-ramp at D16 (Fig. 1 (a)). If the upstream front of synchronized flow of the GP propagates upstream of the upstream bottleneck, then EP emerges which synchronized flow covers these both bottlenecks. Upstream of the bottleneck at D16 a pinch region exists in EP where narrow moving jams emerge and grow (see Fig. 24 in *(7)* and explanations there). However, wide moving jams from the GP downstream of the upstream bottleneck at D16 propagate through this pinch region upstream of D16. These wide moving jams are "foreign" wide moving jams for this pinch region *(7)*.

It occurs that the empirical fundamental for this pinch region shows the same features as for the pinch region of GP (D15, Fig. 8 (b)). This means that the compression of synchronized flow in the pinch region makes more influence on the fundamental diagram than the foreign wide moving jams. However, already 1 km upstream of the pinch region, where the compression of the synchronized flow begins to decrease, the line J is the asymptote for the branch of congested traffic of the empirical fundamental diagram (D14, Fig. 8 (b)). This is although some of the narrow moving jams emerging in the pinch region upstream of the on-ramp do not still transform into wide moving jams. This is linked to the foreign wide moving jam propagation: In the outflow from these jams a flow is formed which is related to averaged points lying on the line J.

Thus, it can be expected that in a lot of cases of the foreign wide moving jam propagation through a congested pattern the branch of congested traffic of the empirical fundamental diagram should asymptotically approach the line J at higher densities in the flow-density plane. This indeed occurs in empirical data where such cases are realized.

## CONCLUSIONS

The investigation made above allows us to conclude the following:

(i) The branch for congested traffic in the empirical fundamental diagram strong depends both on the type of the congested pattern and on the freeway location where the fundamental diagram is measured.

(ii) At freeway locations where the average speed in synchronized flow is higher the branch of the fundamental diagram for congested traffic has a positive slope in the flow-density plane as well the slope of the branch for free flow.

(iii) The average positive slope of the branch for congested traffic decreases in the upstream direction when the average synchronized flow speed inside the congested pattern decreases.

(iv) The branch for congested traffic with a positive slope at a downstream location can transform in the upstream direction in the branch for congested traffic which has a maximum in the flow-density plane: At lower density this branch has a positive and at higher density it has negative slope.

(v) At freeway locations, where the average speed in synchronized flow further decreases moving jams emerge in that synchronized flow. In this case, the branch for congested traffic more and more tends to the line J, i.e., it has the negative slope at each density of congested traffic.

(vi) At upstream locations inside the congested pattern where wide moving jams occur the branch for congested traffic is related to the line J in the flow-density plane.

(vii) The line J plays a role of an asymptote for the branch of congested traffic of empirical fundamental diagrams: In the region of a congested pattern where wide moving jams propagate the branch C for congested traffic almost coincides with the line J at higher densities. This result is linked with the physical sense and the definition of the line J: Each average point of measured data which is related to the wide moving jam outflow should lie on the line J in the flow-density plane. This asymptotic behavior of the branch of congested traffic of empirical fundamental diagrams is essentially important for GPs and EPs where wide moving jams occur.

(viii) The foreign wide moving jam propagation through a congested pattern can lead to the branch of congested traffic of the empirical fundamental diagram which asymptotically approaches the line J at higher densities in the flow-density plane.

## ACKNOWLEDGEMENT

I thank Sergey Klenov for his help and the German Ministry of Education and Research for financial support within the BMBF project "DAISY".

## REFERENCES

1. May, A. D. 1990. Traffic Flow Fundamental (Prentice Hall, Inc., New Jersey).
2. Koshi, M., Iwasaki, M., Ohkura, I. 1983. Some Findings and Overview on Vehicle Flow Characteristics. In: *Proceedings of 8th International Symposium on Transportation and Traffic Theory*, edited by V. F. Hurdle et al. (University of Toronto Press, Toronto, Ontario, 1983) p. 403.
3. Hall, F. L., Allen, B. L., Gunter, M. A.1986. Trans. Res. A **20** 197 (1986).
4. Highway Capacity Manual 2000 (National Research Council, Transportation Research Board, Washington, D.C. 2000).
5. Banks J.H. 1989. Freeway Speed-Flow-Density Relationships: More Evidence and Interpretation. Transportation Research Record **1225** 53-60.
6. Banks J.H. 2002. Review of Empirical Research on Congested Traffic Flow. Transportation Research Record **1802** 225-232.
7. Kerner, B. S. 2002. Empirical features of spatial-temporal traffic patterns at highway bottlenecks. *Phys. Rev. E* **65**, 046138.
8. Kerner, B. S. 2002. Empirical features of congested patterns at highway bottlenecks. *Transportation Research Record* **1802,** 145-154.
9. Kerner, B. S. 2002. Three-Phase Traffic Theory and Highway Capacity. cond-mat/0211684, e-print in http://arxiv.org/abs/cond-mat/0211684.

10. Kerner, B. S. 1998. Experimental Features of Self-Organization in Traffic Flow. *Physical Review Letters* **81**, pp. 3797-3800.
11. Kerner, B. S. 1999. Congested traffic flow: Observations and Theory. *Transportation Research Record* **1678,** 160-167.
12. Kerner, B. S., Rehborn, H. 1976. Experimental Properties of Complexity in Traffic Flow. Phys. Rev. E 53, R4275-R4278.
13. Kerner, B. S. 1998. A Theory of Congested Traffic Flow. In: *Proceedings of the 3rd Symposium on Highway Capacity and Level of Service*, edited by R. Rysgaard, Vol. 2 (Road Directorate, Ministry of Transport - Denmark) pp. 621 – 642.
14. Nishinari, K., Treiber, M., Helbing, D. 2002. cond-mat/0212295.

Figure captions

Fig. 1. Schema of a section of the freeway A5-North in Germany (a), an overview of WSP in space and time (b) and the average speed in WSP as functions of time and detector locations (c).

Fig. 2. States of free flow (black points), states of synchronized flow (circles) for the WSP (D24-D17) and for EP (D16) in comparison with the line J (dashed line J). 1-min averaged data.

Fig. 3. Empirical fundamental diagrams (branches F and C) for WSP in Fig. 2 (b, c) at different locations for the left lane in comparison with the line J. Data are averaged in 3 vehicles/km density intervals.

Fig. 4. The same as in Fig. 3 for data averaged over all three freeway lanes.

Fig. 5. Schema of the freeway A5-South in Germany (a), an overview of GP under strong congestion in space and time (b) and states of free flow (black points) and of synchronized flow (circles) (c).

Fig. 6. Empirical fundamental diagrams (branches F and C) for GP in Fig. 5 (b, c) at different locations for the left lane in comparison with the line J. Data are averaged in 3 vehicles/km density intervals.

Fig. 7. The same as in Fig. 6 for data averaged over all three freeway lanes.

Fig. 8. Empirical fundamental diagrams for GP (a) and for EP (b) at different locations for the left lane in comparison with the line J. Data are averaged in 3 vehicles/km density intervals.

Fig. 9. The same as in Fig. 8 for data averaged over all three freeway lanes.

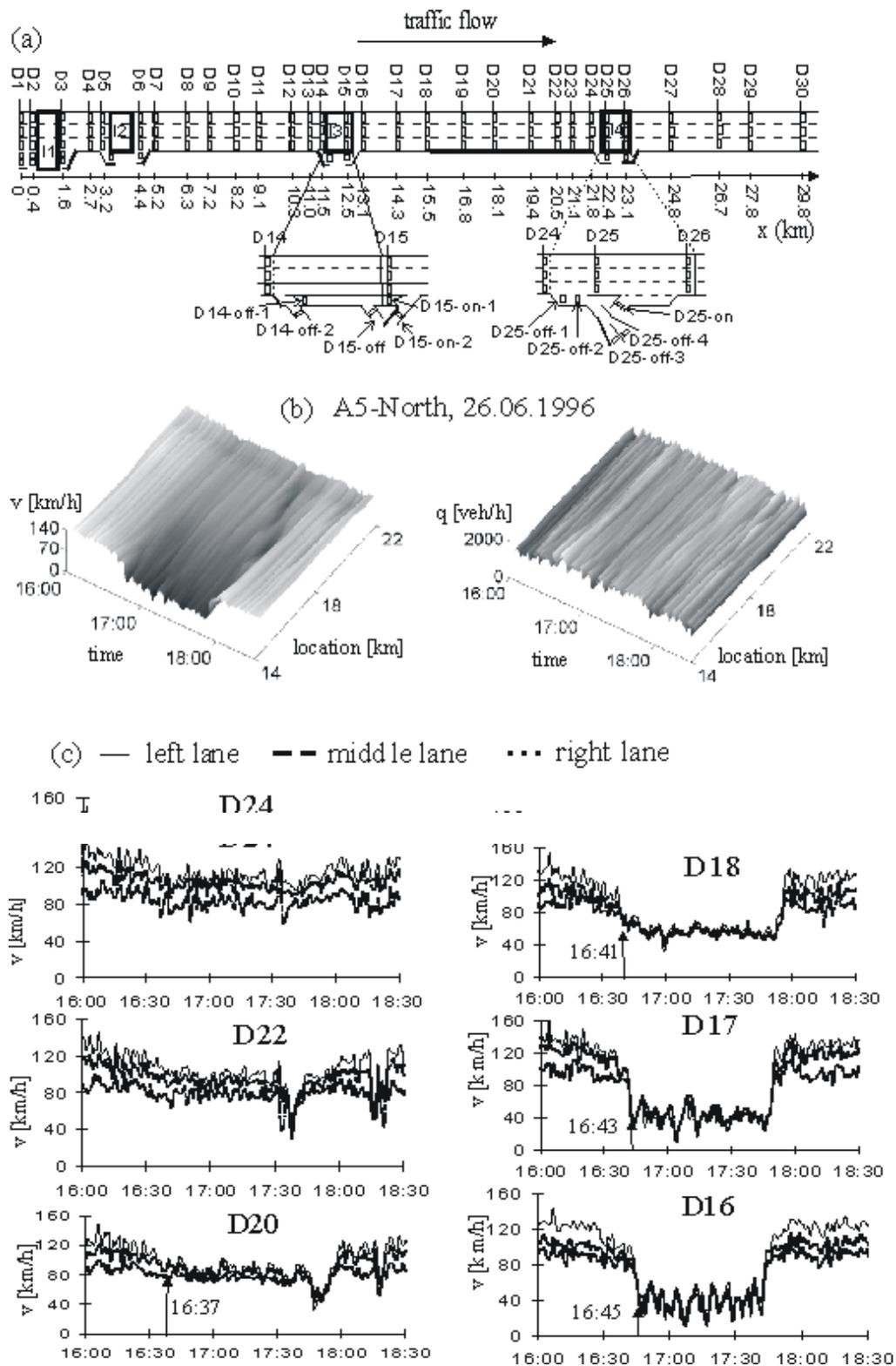

Fig. 1

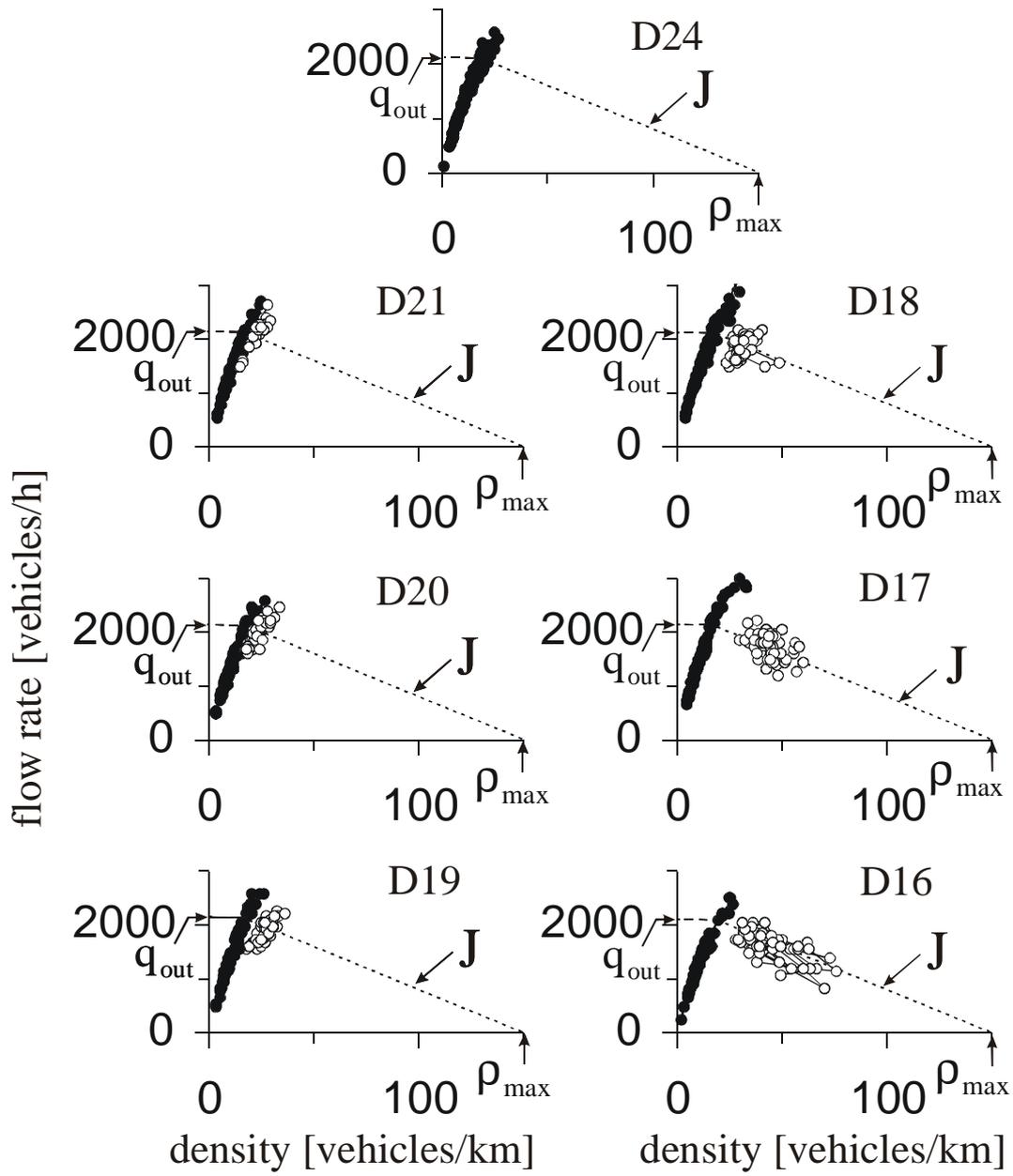

Fig. 2

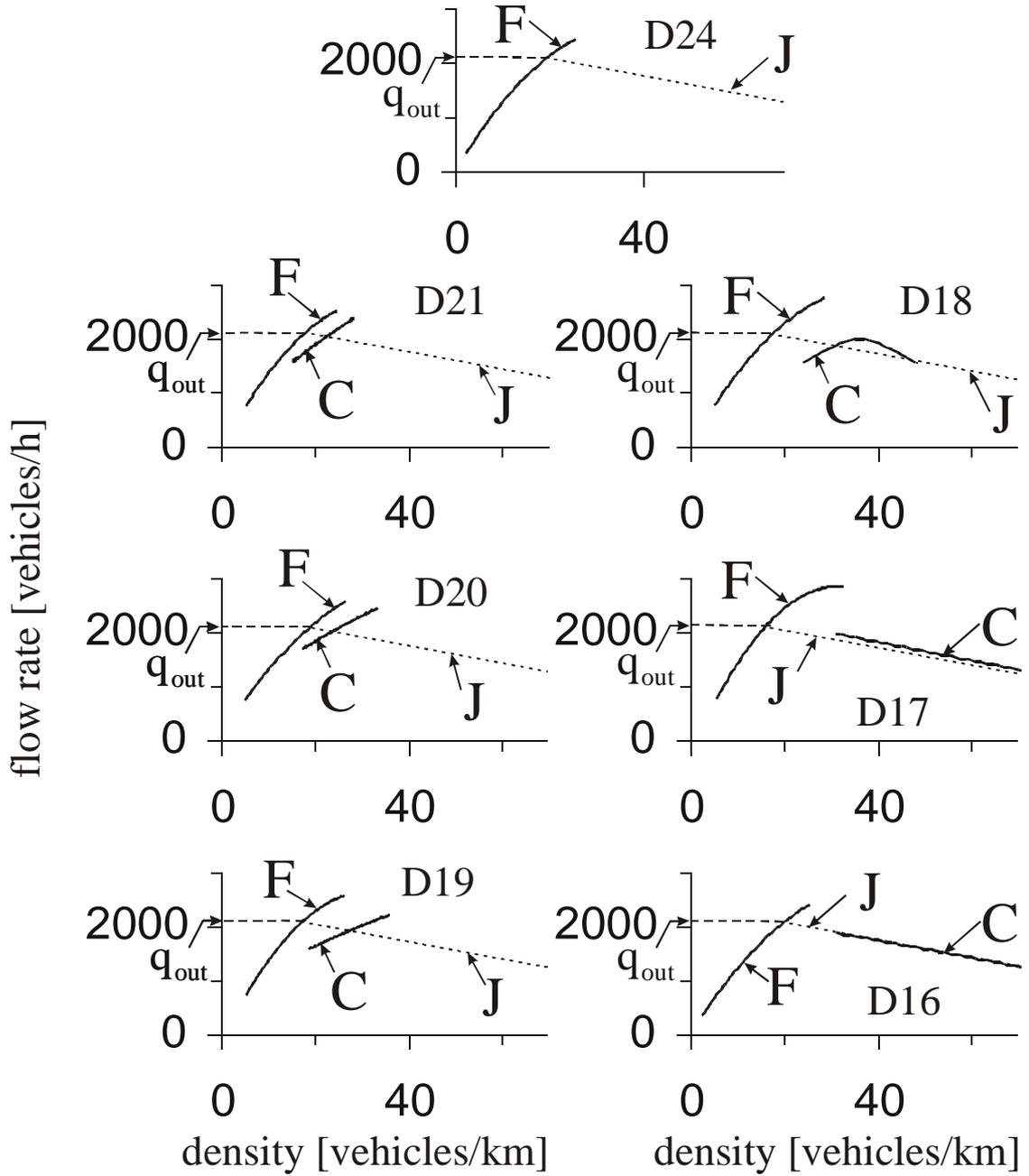

Fig. 3

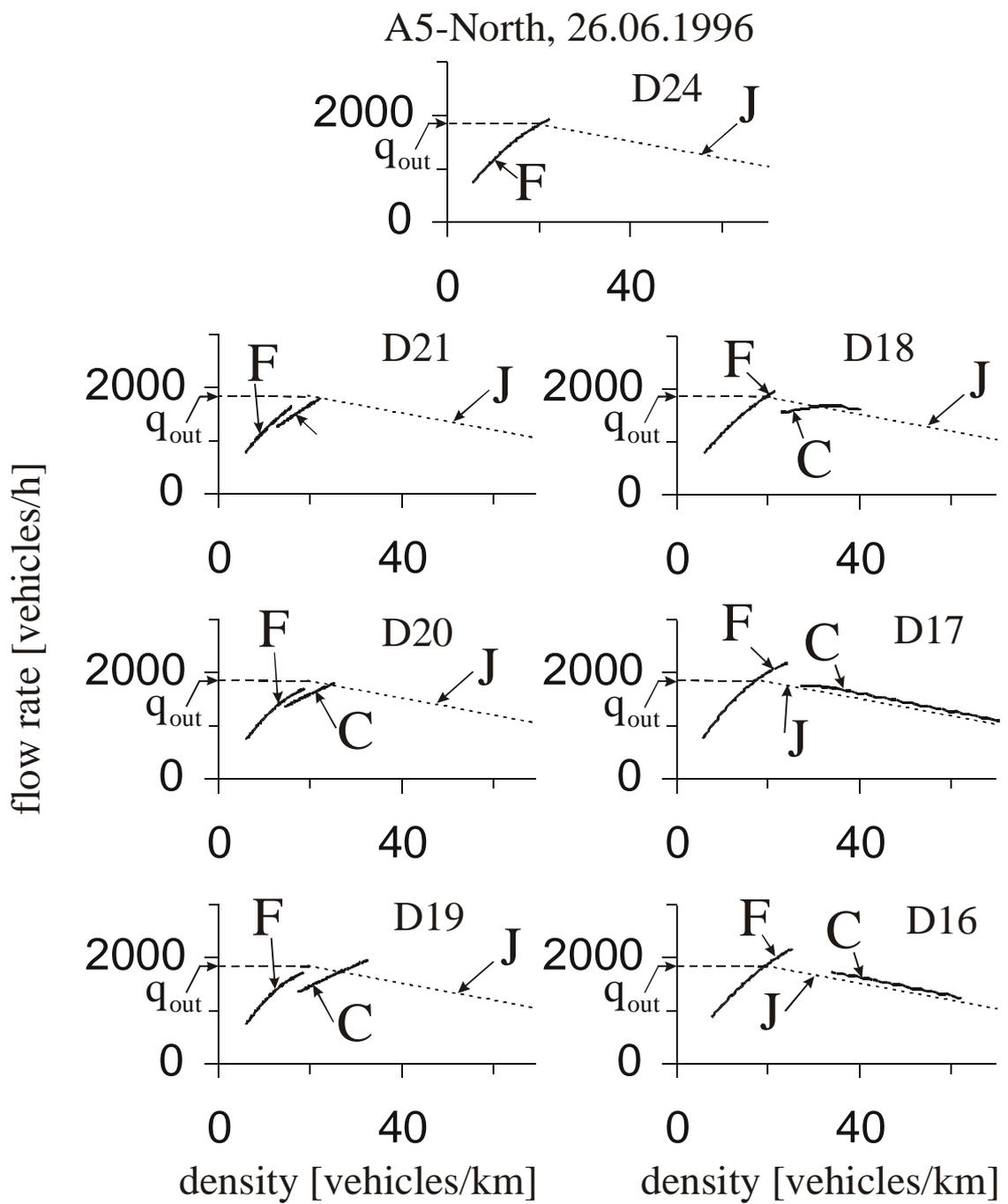

Fig. 4

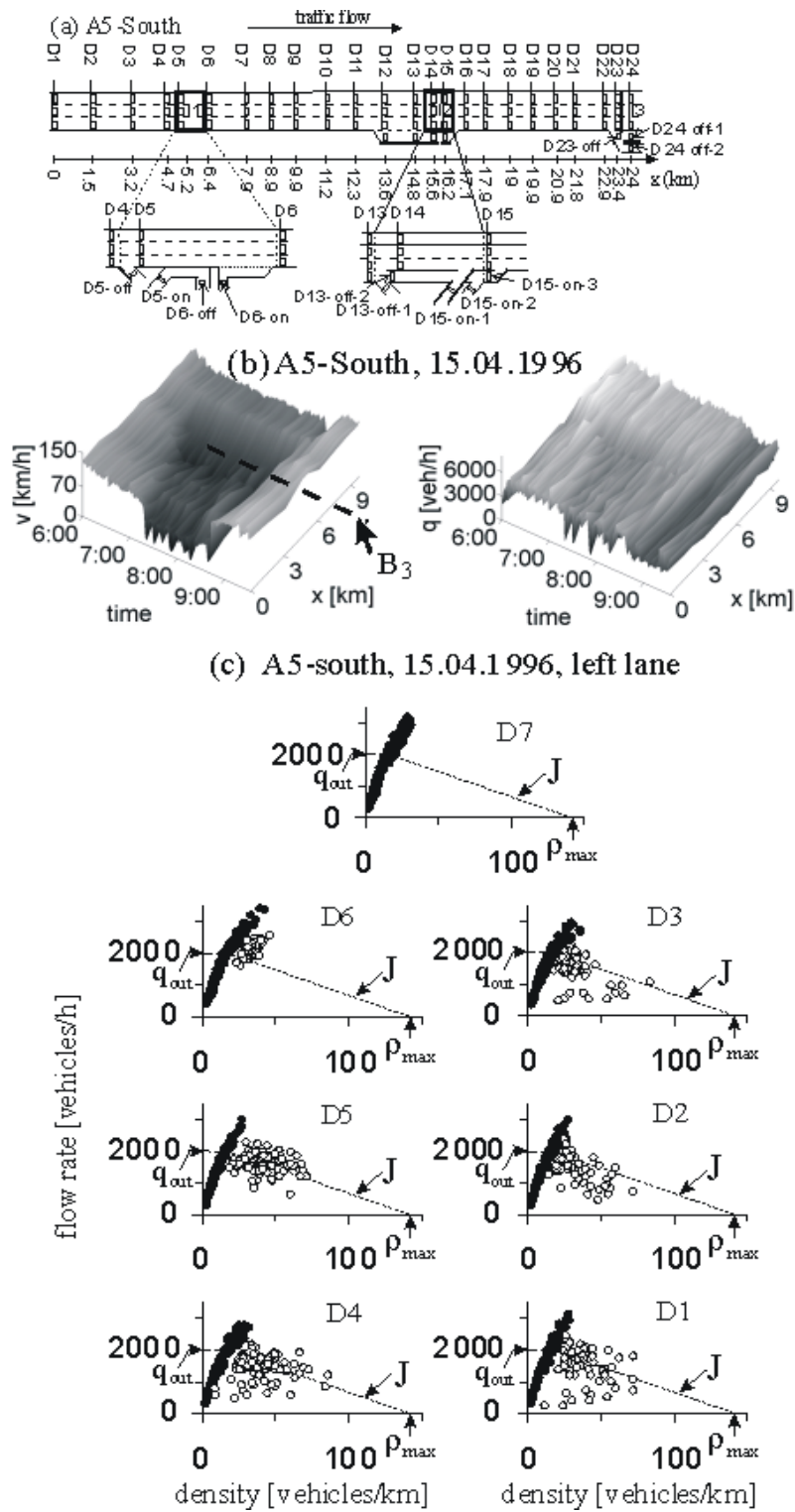

Fig. 5

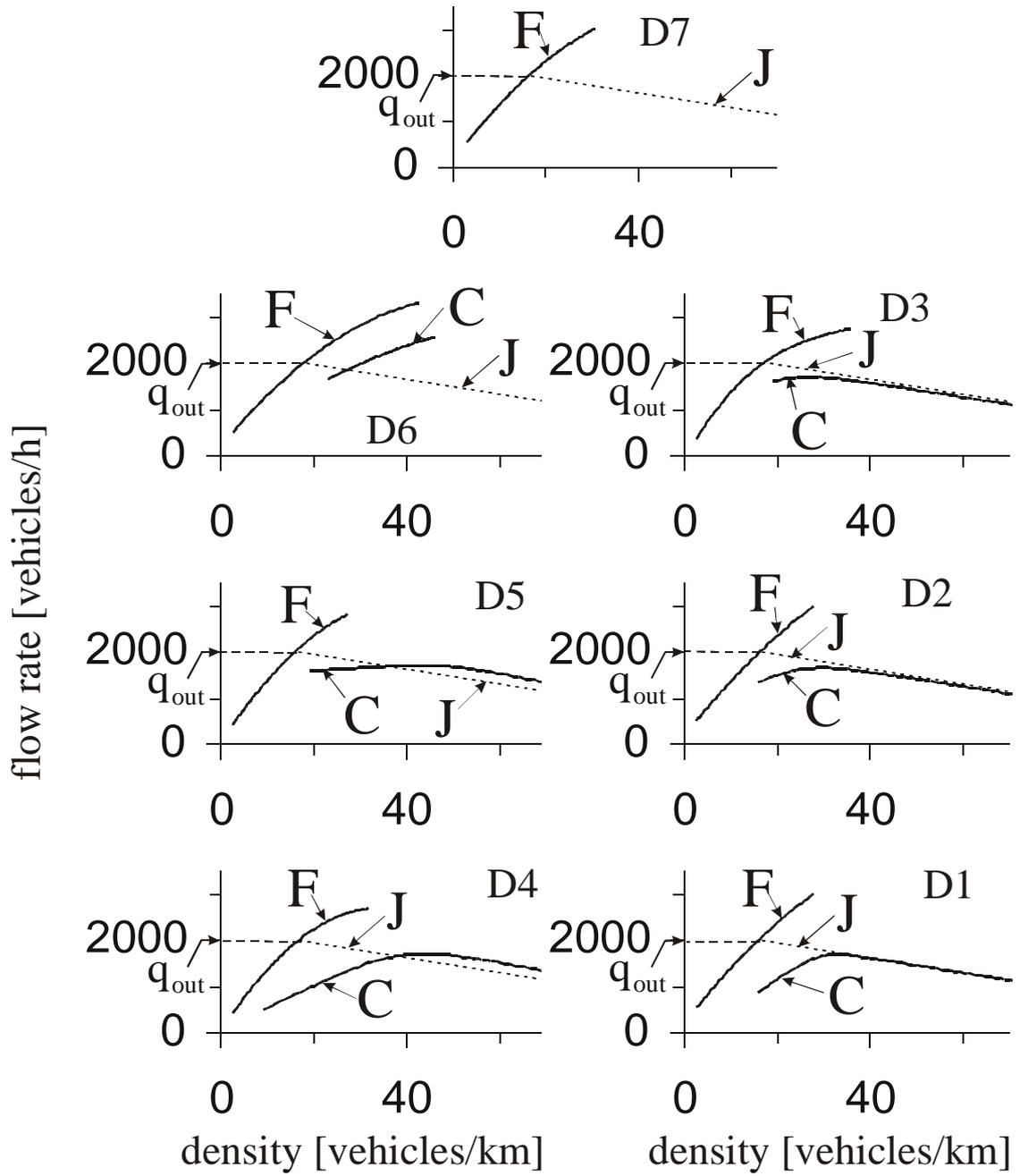

Fig. 6

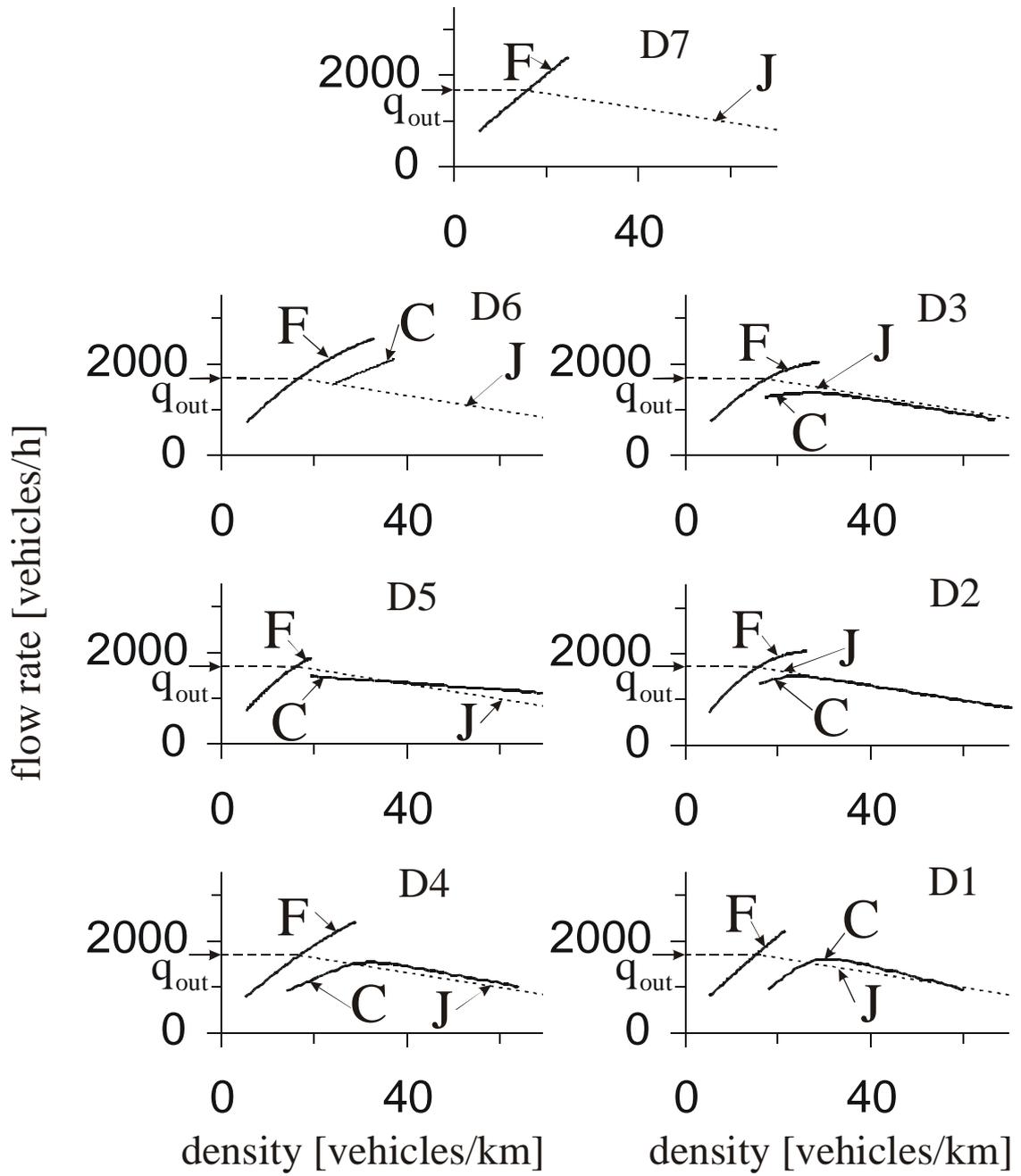

Fig. 7

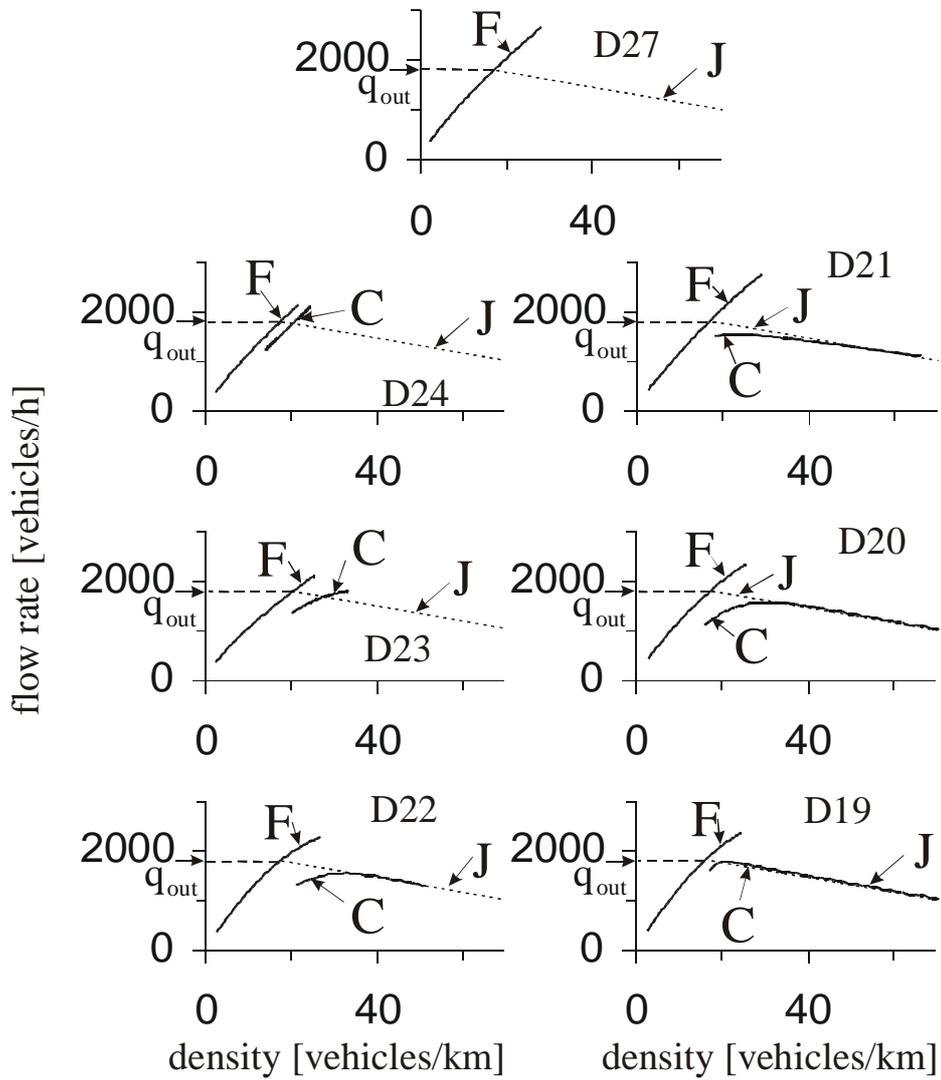

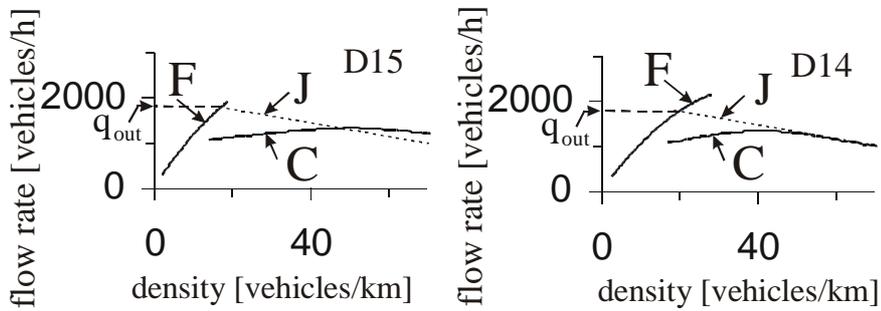

Fig. 8

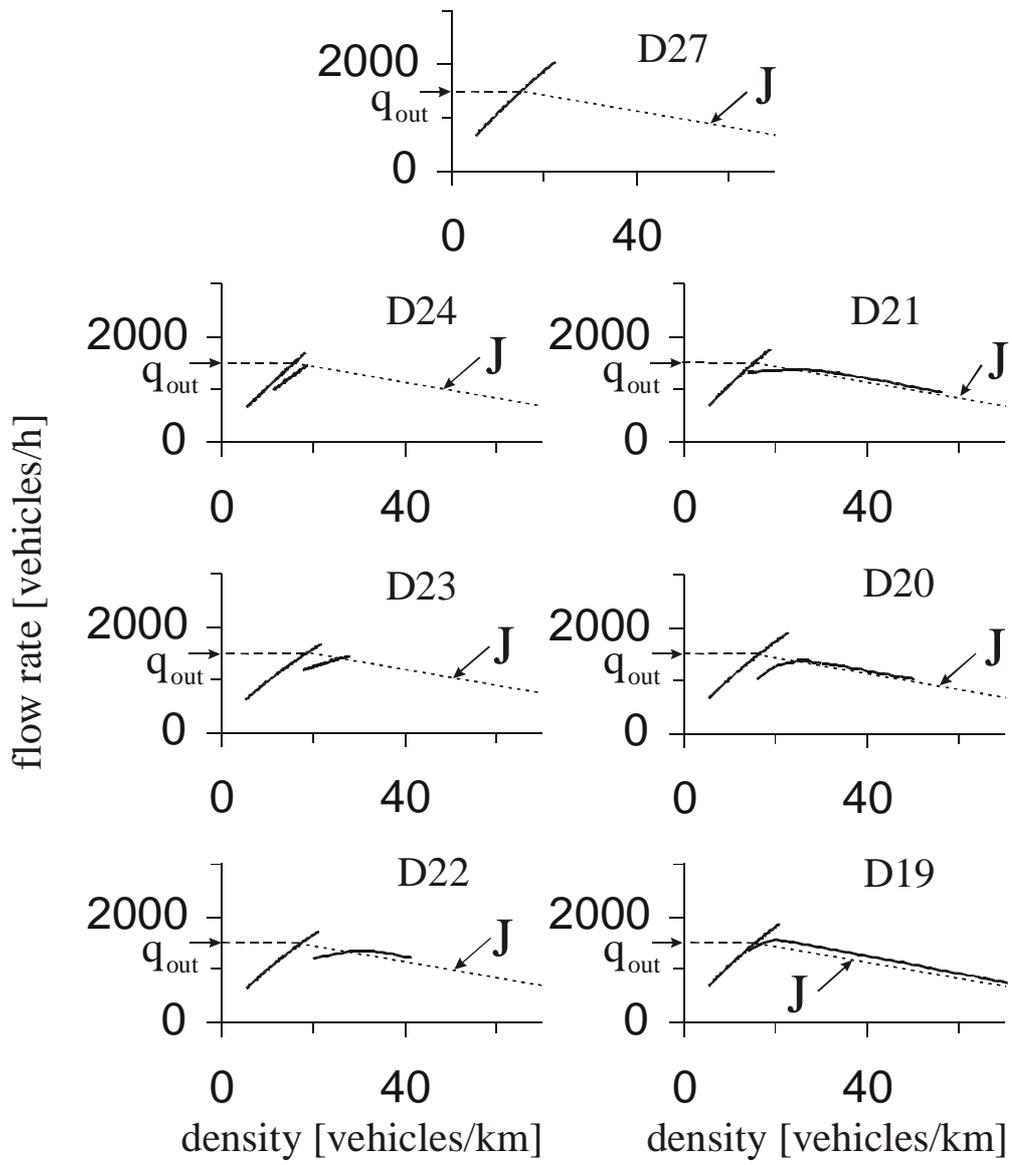

Fig. 9